\documentclass[a4paper,11pt]{article}
\usepackage{pos}

\title{Outlook for the Theoretical Precision of the Luminosity at Future Lepton Colliders}
\centerline{BU-HEPP-24-02, 09/2024}
\author*[a]{B.F.L. Ward}
\author[b]{S. Jadach\footnote[2]{Deceased.}} 
\author[c]{W. Placzek}
\author[b]{M. Skrzypek}
\author[d]{S.A. Yost}

\affiliation[a]{Department of Physics, Baylor University,\\
  Waco, TX, USA}

\affiliation[b]{Institute of Nuclear Physics,\\
Krakow, PL}

\affiliation[c]{Institute of Applied Computer Science, Jagiellonian University,\\
Krakow, PL}

\affiliation[d]{Department of Physics, The Citadel,\\
  Charleston, SC, USA}

\emailAdd{bfl\_ward@baylor.edu}
\emailAdd{wieslaw.placzek@uj.edu.pl}
\emailAdd{maciej.skrzypek@ifj.edu.pl}
\emailAdd{yosts1@citadel.edu}

\abstract{The LEP precision physics requirements on the theoretical precision tag for the respective luminosity were  $0.054 \%$ ($0.061\%$) at $M_Z$, where the former (latter) LEP result has (does not have) the pairs correction. For the contemplated FCC-ee, ILC, and CEPC Higgs/EW factories, one needs improvement at $M_Z$ to at least $0.01\%$ for the theoretical precision tag. We discuss the paths one may take to even exceed this latter goal and present an update on the current expectations for both $M_Z$ and proposed higher energy scenarios.}

\FullConference{42nd International Conference on High Energy Physics (ICHEP2024)\\
18-24 July 2024\\
Prague, Czech Republic\\}

\begin{document}
\maketitle

\section{In Memoriam to Prof. Stanislaw Jadach}
\begin{figure}[h]
\begin{center}
\setlength{\unitlength}{1in}
\begin{picture}(6,4.5)(0,0)
\put(0.5,1.5){\includegraphics[width=5in]{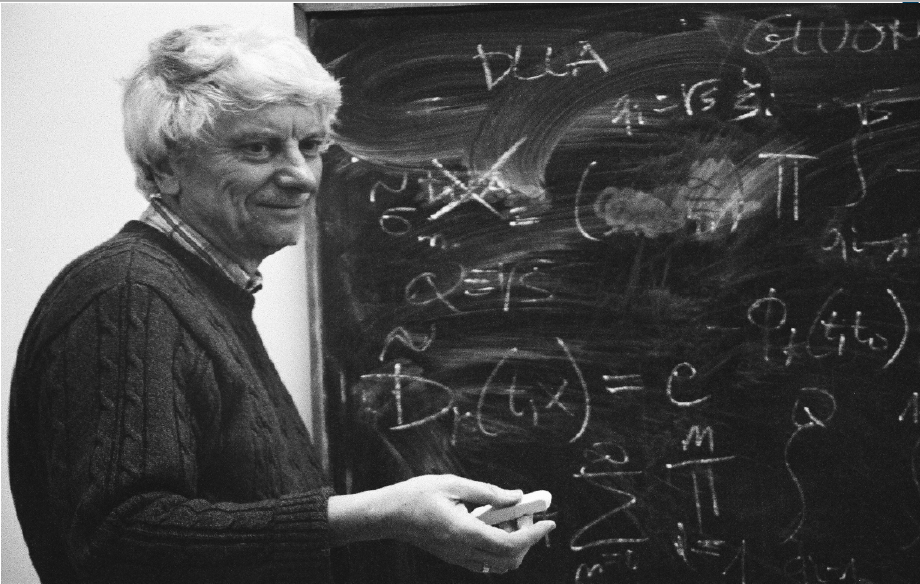}}
\end{picture}
\end{center}
\vspace{-42.5mm}
\caption{The late Prof. Stanislaw Jadach. See Ref.~\cite{cern-courier-2023} for his {\it CERN Courier} obituary.}
\label{fig1}
\end{figure}

This poster is dedicated in memoriam to our close friend and colleague Prof. Stanislaw Jadach (shown here in Fig.~\ref{fig1}) who passed away suddenly on Feb. 26, 2023~\cite{cern-courier-2023}. We miss him dearly. As this is the first Rochester Conference since his passing, we lift up his special contributions to the understanding of the lumi theory error problem at LEP/SLC and at possible future lepton colliders.
\section{Current Theoretical Luminosity Precision Outlook}
\begin{figure}[htp!]
\begin{center}
\setlength{\unitlength}{1in}
\begin{picture}(6,3.5)(0,0)
\put(0.5,1.5){\includegraphics[width=5in]{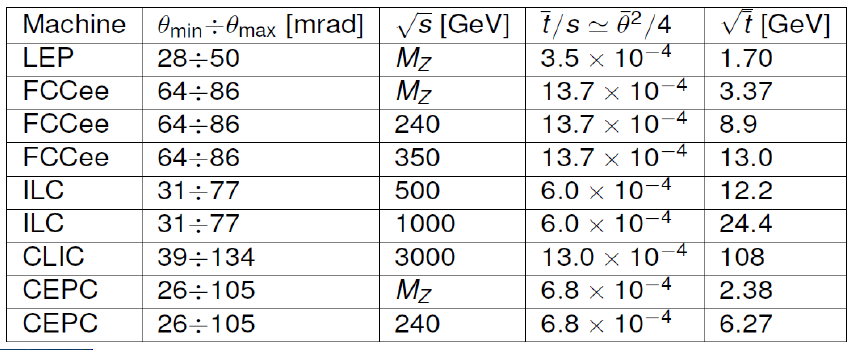}}
\end{picture}
\end{center}
\vspace{-42.5mm}
\caption{Comparison of detector acceptances for previous and possible future lepton colliders.}
\label{fig2}
\end{figure}

It is well known that the Bhabha cross-section depends on the detector's acceptance as we illustrate with the formula
\begin{equation}
\sigma_{\cal L}\simeq 4\pi \alpha^2\Large(\frac{1}{|t|_{min}} - \frac{1}{|t|_{max}}\Large)=4\pi \alpha^2\Large(\frac{|t|_{max}-|t|_{min}}{\bar{t}^2}\Large),\; \bar{t}\equiv \sqrt{|t|_{min}|t|_{max}}.
\label{eq1}
\end{equation}
We see that the characteristic scale of the process $\bar{t}$ gives $\bar{t}/s$ as the suppression factor between s- and t-channel contributions. The comparison of the  latter factor for previous and possible future lepton colliders is shown in Fig.~\ref{fig2}. The t-channel dominance factor is generically at the per-mille level.\par
\begin{figure}[h!]
\begin{center}
\setlength{\unitlength}{1in}
\begin{picture}(6,3.5)(0,0)
\put(0.0,2.4){\includegraphics[width=3in,height=1.15in]{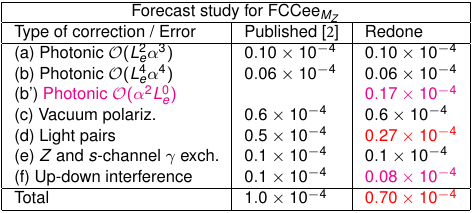}}
\put(3,2.4){\includegraphics[width=3in,height=1.17in]{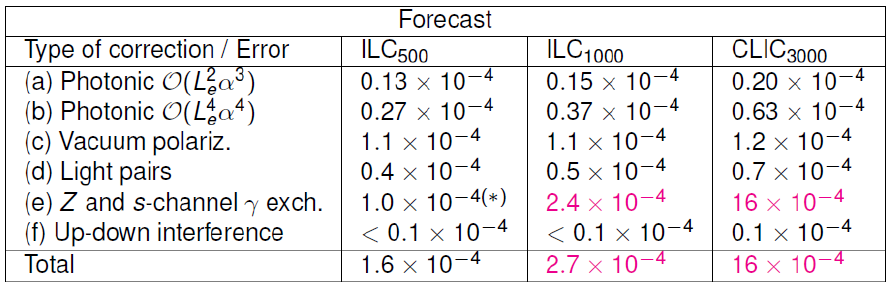}}
\end{picture}
\end{center}
\vspace{-65mm}
\caption{On the left, lumi theory outlook when redone as in Refs.~\cite{fcc2023wkshpms,skrzypek-2023mttd} -- lattice~\cite{latt1,latt2,latt3} treatment of (c) could reduce it by $\sim 6$. On the right, expectations for higher energies.}
\label{fig3}
\end{figure}

In Refs.~\cite{Jadach:2018jjo,Jadach:2021ayv,fcc2023wkshpms,skrzypek-2023mttd}, exploiting the suppression between s- and t-channel contributions, we have shown that we have a path to 0.01\% precision for our BHLUMI~\cite{bhlumi4:1996} MC and, more recently, that the path to 0.001\% is not closed. This is illustrated in Fig.~\ref{fig3}, wherein we also show expectations for the higher energies. The key issue is the availability of funding for the required resources. Presumably, once the decision to move forward with a specific Higgs/EW Factory option is more certain, the attendant theory support will obtain.\par
\section*{Acknowlegements} S.A. Yost was supported in part by a grant from the Citadel Foundation. The authors thank Prof. G. Giudice for the support and kind hospitality of the CERN TH Department. 
S.J. acknowledges funding from the European Union’s Horizon 2020 research and innovation programme 
under under grant agreement No 951754 and support of the National Science Centre, Poland, Grant No. 
2019/34/E/ST2/00457.
\baselineskip=10pt
\bibliography{Tauola_interface_design}{}

\providecommand{\href}[2]{#2}\begingroup\begin{thebibliography}{1}

\bibitem{cern-courier-2023}
W.~P\l{}aczek, M.~Skrzypek, B.~F.~L. Ward, and Z.~W\c{a}s, {\em CERN Courier}
  {\bf 63} (2023), no.~3, 59.

\bibitem{Jadach:2018jjo}
S.~Jadach, W.~P\l{}aczek, M.~Skrzypek, B.~F.~L. Ward, and S.~A. Yost, {\em
  Phys. Lett. B} {\bf 790} (2019) 314--321,
  \href{http://www.arXiv.org/abs/1812.01004}{{\tt 1812.01004}}.

\bibitem{Jadach:2021ayv}
S.~Jadach, W.~P\l{}aczek, M.~Skrzypek, and B.~F.~L. Ward, {\em Eur. Phys. J. C}
  {\bf 81} (2021), no.~11 1047.

\bibitem{fcc2023wkshpms}
M.~Skrzypek {\em et al.}, talk in {\it 2023 FCC Workshop, Krakow, PL}.

\bibitem{skrzypek-2023mttd}
M.~Skrzypek, W.~P\l{}aczek, B.~F. Ward, and S.~Yost, ``{How well could we
  calculate luminosity at FCCee?}'', in {\em {Proc. XXX MTTD Conference,
  Ustro\'n, PL, 2023}}, vol.~XXX, in press.

\bibitem{bhlumi4:1996}
S.~Jadach, W.~Placzek, E.~Richter-W\c{a}s, B.~F.~L. Ward, and Z.~W\c{a}s, {\em
  Comput. Phys. Commun.} {\bf 102} (1997)
229.

\bibitem{latt1}
S.~Borsanyi {\em et al.},
  \href{http://www.arXiv.org/abs/hep-lat/1711.04980}{{\tt hep-lat/1711.04980}}.

\bibitem{latt2}
M.~Ce {\em et al.}, \href{http://www.arXiv.org/abs/hep-lat/2203.08676}{{\tt
  hep-lat/2203.08676}}.

\bibitem{latt3}
Z.~Fodor, talk in {\it ICHEP2024}.

\end{thebibliography}\endgroup
\bibliographystyle{utphys_spires}

\end{document}